\documentclass[a4paper,fleqn,usenatbib]{mnras}

\usepackage[T1]{fontenc}
\usepackage{ae,aecompl}
\usepackage{hyperref}

\usepackage{graphicx}	
\usepackage{amsmath}	
\usepackage{amssymb}	

\title[A new region of $\gamma$-ray emission near HESS J1825-137]{Revealing a new region of gamma-ray emission in the vicinity of HESS J1825-137}

\author[Araya et al.]{
M. Araya,$^{1}$\thanks{E-mail: miguel.araya@ucr.ac.cr}
A.M.W. Mitchell,$^{2,3}$
R.D. Parsons,$^{3}$
\\
$^{1}$Centro de Investigaciones Espaciales and Escuela de F\'isica, Universidad de Costa Rica\\
$^{2}$Physik Institut, Universit\"{a}t Z\"{u}rich, Winterthurerstrasse 190, CH-8057 Z\"{u}rich, Switzerland\\
$^{3}$Max-Planck-Institut f\"{u}r Kernphysik, P.O. Box 103980, D-69029 Heidelberg, Germany}

\date{Accepted ---. Received ---; in original form ---}

\pubyear{2018}

\begin{document}
\label{firstpage}
\pagerange{\pageref{firstpage}--\pageref{lastpage}}
\maketitle

\begin{abstract}
HESS J1825-137 is a bright very-high-energy (VHE) gamma-ray source that has been firmly established as a pulsar wind nebula (PWN), and one of the most extended gamma-ray objects within this category. The progenitor supernova remnant (SNR) for this PWN has not been firmly established. We carried out an analysis of gamma-ray observations in the region of HESS J1825-137 with the \emph{Fermi}-LAT which reveal emission in the direction away from the Galactic plane. The region lies beyond the PWN and reaches a distance from the pulsar compatible with the supposed location of the SNR shock front. The spectrum of the gamma-rays is hard with a photon index of $\sim 1.9$ in the 10-250 GeV range. Several scenarios for the origin of the emission are discussed, including the SNR as a source of high-energy particles and the ``leakage'' of leptons from the PWN.
\end{abstract}

  
\section{Introduction} \label{sec:intro}
Pulsar wind nebulae (PWN) are produced by the flow of a relativistic pair plasma and electromagnetic fields generated by spin-powered pulsars. These particles emit synchrotron radiation from radio to X-rays or even higher energies and also show an inverse Compton (IC) component at gamma-ray energies \citep[e.g.,][]{2006ARA&A..44...17G,2015SSRv..191..391K}. PWN are sometimes seen to expand within the associated supernova remnant (SNR), which are also known for accelerating particles \citep[e.g.,][]{2016ApJS..224....8A}.

HESS J1825-137 is one of the most luminous of the identified PWN. It was classified as a PWN on the basis of its energy dependent morphology at TeV energies \citep{2006A&A...460..365A}. It is powered by the pulsar PSR J1826-1334, which has a spin-down power of $\dot E = 2.8\times 10^{36}$ erg/s and a characteristic age of 21 kyr \citep{1992MNRAS.254..177C}. Its dispersion measure distance is estimated to be $3.9\pm 0.4$ kpc \citep{1993ApJ...411..674T,2005AJ....129.1993M,2002astro.ph..7156C}. The PWN extends towards the south of the pulsar location. The interaction of the progenitor SNR with molecular clouds north of the pulsar has been suggested as the plausible reason for this asymmetry \citep[e.g.,][]{2008ApJ...675..683P}. X-ray observations show the PWN emission extending at least 17 pc south of the pulsar. High-energy electrons continued synchrotron emission at such a large distance requires either diffusion parallel to the magnetic field with a diffusion coefficient of $\sim 5\times 10^{28}$ cm$^2$ s$^{-1}$ or convective processes \citep{2009PASJ...61S.189U}.

A large rim of H$\alpha$ emission with spectroscopic signatures suggestive of shocked gas from an SNR was found by \cite{2008MNRAS.390.1037S}. If located at the same distance as the pulsar powering HESS J1825-137, the radius of the shock front would be $\sim$120 pc. At a similar angular distance, and located to the south-east of the pulsar, another rim-like enhancement of H$\alpha$ emission has been noted by \cite{2016MNRAS.458.2813V} in their interstellar medium (ISM) studies of the region, which could then be caused by the same SNR shock.

The large radius of the candidate host SNR as well as the anomalously large extent of the TeV emission from the PWN could be explained if the pulsar had a relatively large initial spin-down power and the SNR expanded into hot ISM with very low density \citep[$n \sim 0.003$ cm$^{-3}$,][]{2009ASSL..357..451D}. However, typical ISM densities are much greater than this value and observations of molecular clouds in the region of HESS J1825-137 would also challenge this assumption. Recently, calculations by \cite{2018ApJ...860...59K} showed that the low ISM density requirement could be relaxed if the pulsar has a small braking index ($\leq 2$) and the pulsar birth period was short ($\sim 1$ ms).

Here we present an analysis of gamma-ray data from the \emph{Fermi} Large Area Telescope (LAT), a pair conversion telescope that detects photons in the energy range between 20 MeV and $> 500$ GeV \citep{2009ApJ...697.1071A}, in the region of the SNR possibly associated with HESS J1825-137. The data reveal substantial emission extending several degrees and partly sorrounding the TeV emission from the PWN. We also study the properties of this region using publicly available multiwavelength observations. The spectral properties and energy content of the high-energy radiation are derived and we discuss different possibilities to explain its origin.

\section{\emph{Fermi} LAT data} \label{sec:LAT}
We analyzed data taken between MJD 54682 (2008 August 4) and MJD 58289 (2018 June 20) with the Pass8 response functions and the latest version (v11r5p3) of the ScienceTools. We selected time intervals with good data quality, as well as events having a maximum zenith angle of 90$\degr$ to avoid contamination by gamma-rays produced in Earth's limb, reconstructed within a radius of 20$\degr$ around a position to the south of HESS J1825-137 at the coordinates RA=$18^h30^m00^s$, Dec=$-15\degr30'00''$ (J2000). In order to take advantage of the improved LAT resolution at higher energies, we used events above an energy threshold of 10 GeV for which the 68\%-containment point-spread function (PSF) is $< 0.2 \degr$. LAT data are divided into classes depending on the photon probability and reconstruction quality and each class is subdivided into event types. For instance, some event types are used to separate events that are converted in the front or back section of the tracker\footnote{\url{https://fermi.gsfc.nasa.gov/ssc/data/analysis/documentation/Cicerone/Cicerone\_Data/LAT\_DP.html}}. For most analyses we used combined front and back events in the P8R2\_SOURCE class to increase the statistics and for the spatial profiles (see below) we used front events only in the P8R2\_CLEAN class to lower the background rate at the energies involved, improve the PSF ($\leq 0.1 \degr$, 68\%-containment) and slightly increase the sensitivity for hard sources\footnote{\url{http://www.slac.stanford.edu/exp/glast/groups/canda/lat\_Performance.htm}}.

The LAT data were fitted with a model including the gamma-ray sources in the region found in the Third Catalog of Hard Fermi-LAT Sources \citep[detected above 10 GeV,][hereafter 3FHL]{2017ApJS..232...18A}. The model included the Galactic diffuse emission (described by the file gll\_iem\_v06.fits) and the residual background and extragalactic emission (contained in the file iso\_P8R2\_SOURCE\_V6\_v06.txt, or the corresponding variation for a different event class and type when used). The maximum likelihood method \citep{1996ApJ...461..396M} was used to fit the data. During the fit we left the spectral normalizations of the sources located within $10\degr$ from the center of the region free to vary and kept the spectral shape parameters fixed. The exceptions were the sources 3FHL J1826.2-1451 (associated to LS 5039) and 3FHL J1823.3-1339 to the north of the PWN. For these two sources all the spectral parameters were left free since the fit was producing substantial negative residuals at their locations. The spectrum of LS 5039 is known to be variable which could explain the over-subtraction. In the case of 3FHL J1823.3-1339 the fit could be affected by the bright emission from the PWN itself. We also kept fixed all the parameters of the sources located more than $10\degr$ away from the center of the region. The unidentified point source 3FHL J1830.0-1518, reportedly having a hard spectral index ($\sim 1.7$) and located $\sim 2\degr$ to the south of PSR J1826-1334, was removed from the model in order to carry out a more detailed analysis of the emission. Additional extended templates were added to the model to account for emission in the regions of the TeV sources HESS J1809-193 and HESS J1813-178, as found by \cite{2018ApJ...859...69A}, although these additions had no effect on the results described here. All error bars reported in this paper were calculated at the $1\sigma$ level unless otherwise noted.

\begin{figure}
  \includegraphics[width=9.7cm,height=10cm]{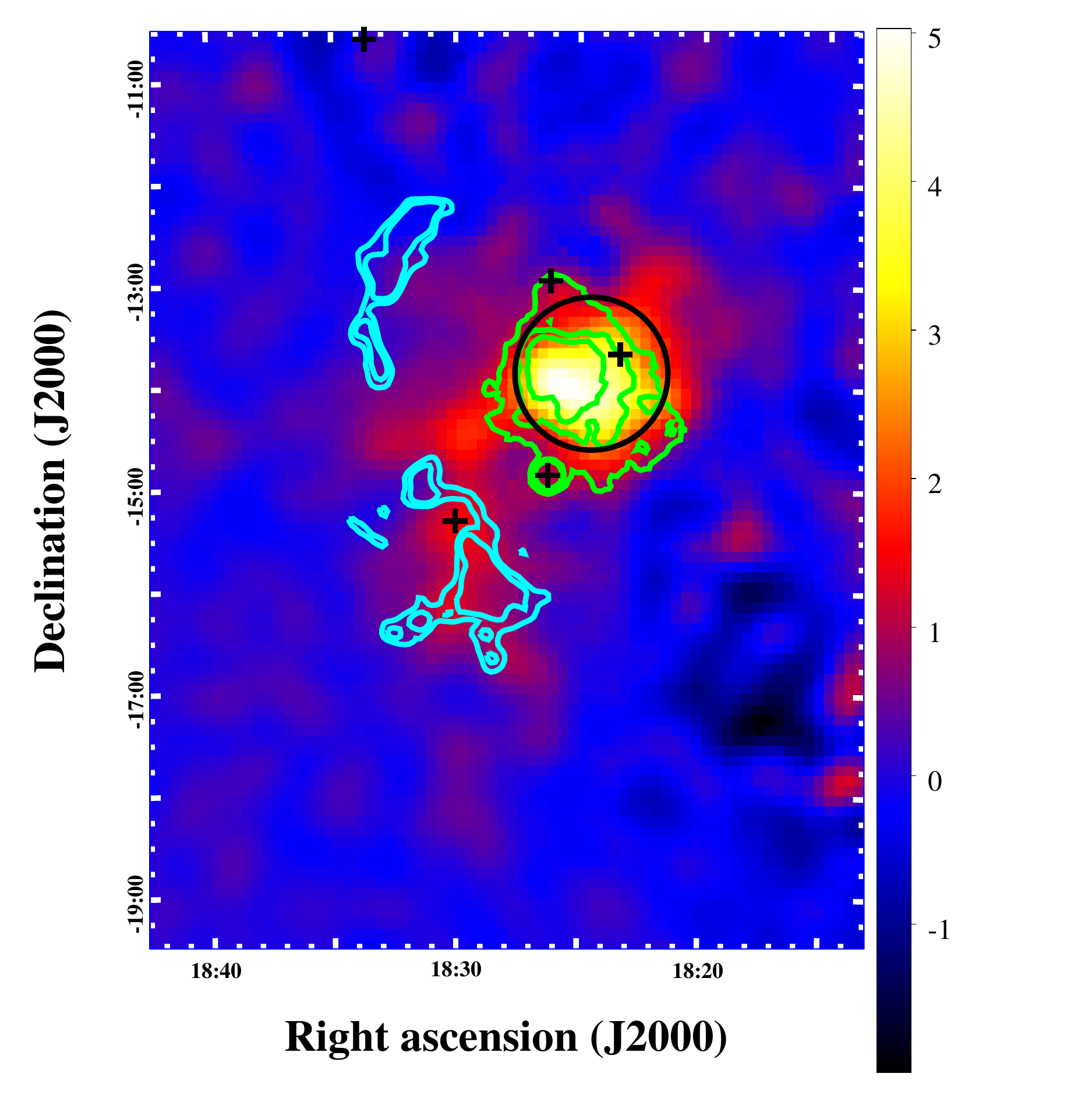}
  \caption{Residual count map obtained from LAT data above 10 GeV in the region of HESS J1825-137. The map was smoothed by a Gaussian kernel with $\sigma = 0.2\degr$ and the pixel size is $0.1\degr$. The cyan contours represent the H$\alpha$ emission thought to be associated with the host SNR shock front (at the levels of 90, 112, 240, 960 and 5000 Rayleighs), while the green contours are the TeV emission at the 5, 10 and 15$\sigma$ levels. The black crosses and circle are the 3FHL sources in the region.}
  \label{fig1}
\end{figure}

After subtracting the best-fit model from the data counts map the resulting residuals are shown in Fig. \ref{fig1}. The LAT template for HESS J1825-137 was not included in the model to obtain the map and this source can be seen as well. The TeV emission of the PWN \citep[taken from the publicly available $0.1\degr$-correlation radius map of the H.E.S.S. Galactic plane survey][]{2018A&A...612A...1H} is outlined by the green contours while the H$\alpha$ emission can be seen as cyan contours \citep[taken from][]{2003ApJS..146..407F}. The H$\alpha$ emission to the left of the image is believed to be associated with shocked gas tracing the SNR shock front. Substantial, apparently extended, GeV emission is seen covering a large portion of the image and its properties are studied in the following sections.

\subsection{Morphology of the emission}\label{morphology}
\begin{figure}
  \includegraphics[width=9.7cm,height=10cm]{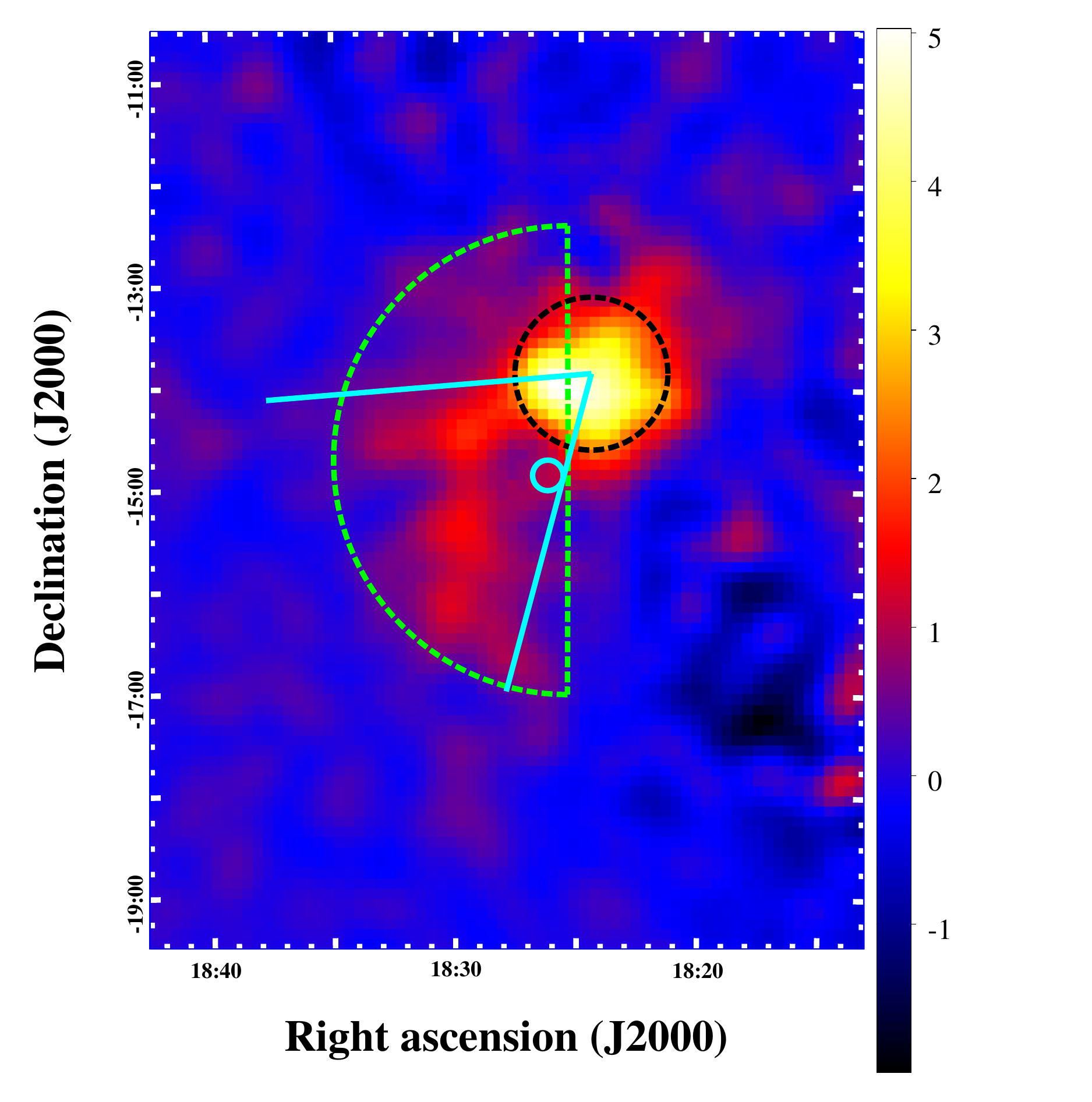}
  \caption{LAT residual count map as shown in Fig. \ref{fig1}. The dashed semicircle shows the border of the best-fit template found in this work and the solid lines the wedge used for obtaining the radial brightness profile. The black dashed circle indicates the 68\% containment radius ($0.75\degr$) of the 2D Gaussian representing HESS J1825-137 in the 3FHL catalog. The smaller circle represents the exclusion region for the point source LS 5039.}
  \label{fig2}
\end{figure}

Since the residual GeV emission resembles a circular sector extending mainly from the north-east to the south-east of the PWN, we aimed to quantify its flux and spectral properties under the extended hypothesis by using a simple geometrical template. The adopted morphology was a semicircle of uniform brightness. We fitted the location of the center (which is defined as the centre of the full circle described by the same radius), inclination and radius of the template to maximize the test statistic (TS), defined as $-2\times$ln$(L_0/L)$, with $L$ and $L_0$ the maximum values of the likelihood functions for models with the template and the null hypothesis (no template for the emission), respectively. For all the fits the standard template representing HESS J1825-137 was kept in the model. The resulting center of the best-fit semicircle, with TS = 281.5, was found at the coordinates RA=$18^h25^m30^s$, Dec=$-14\degr42'08''$ (J2000), and the optimized radius found was $2.3^{+0.3}_{-0.1}\degr$. From the change in TS for different locations of the template we estimated an uncertainty of $0.2\degr$ ($3\sigma$ level) for the center position. The orientation of the diameter of the best-fit template follows a line of constant RA=$18^h25^m30^s$ as shown in Fig. \ref{fig2}. We also found that for templates covering more than half of the region occupied by the bulk of the GeV gamma-rays from HESS J1825-137 the TS values started to become slightly larger, and we believe this was caused by the influence of leftover emission around the PWN. We leave a more detailed analysis of the region for a future publication.

\begin{figure}
  \includegraphics[width=9.7cm,height=9.7cm]{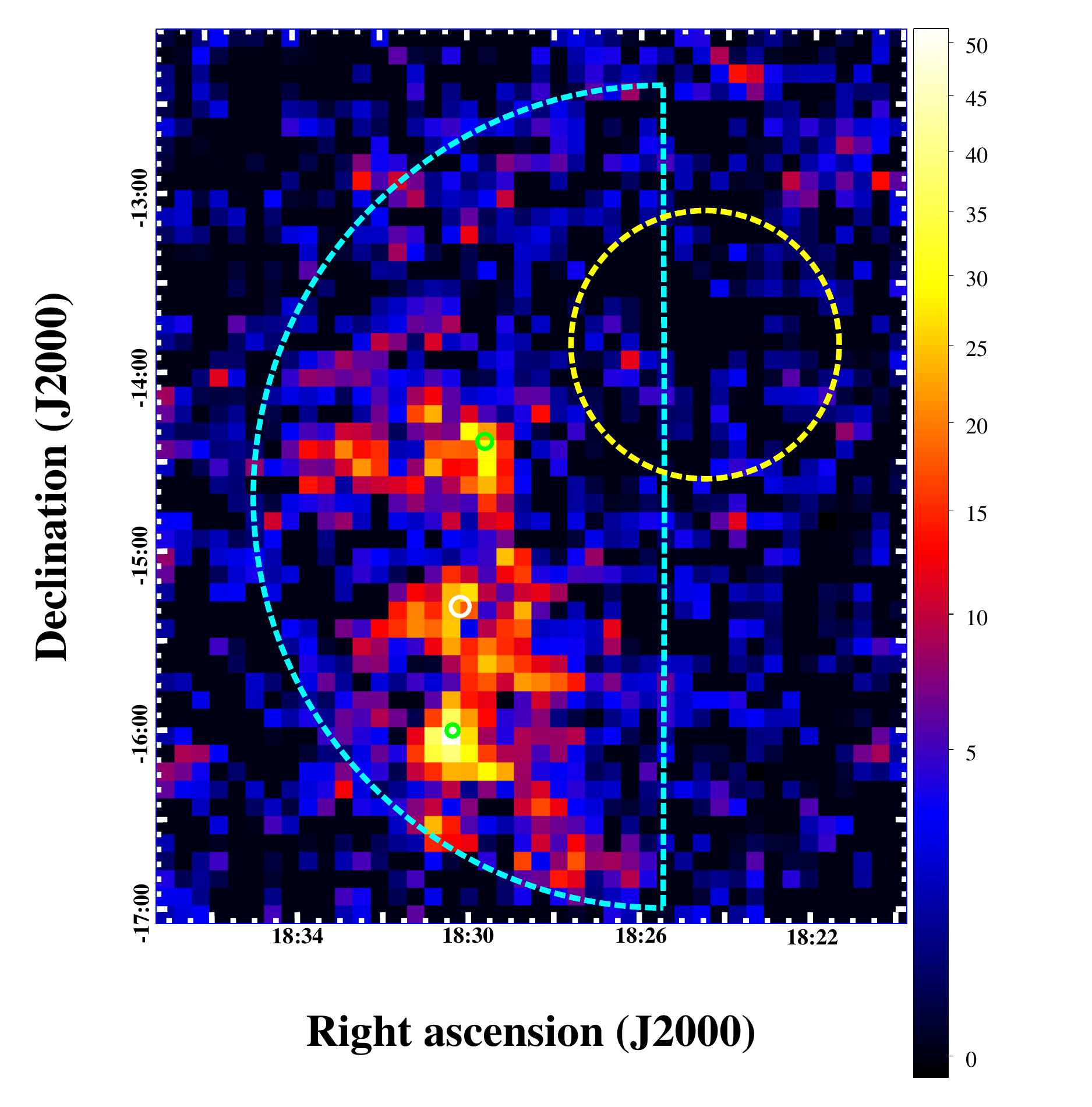}
  \caption{TS map for a point source hypothesis in the region around HESS J1825-137, which is shown as a yellow circle. The template for the new emission shown in Fig. \ref{fig2} is also seen. The small green circles represent the two most significant of the sources found in the point source search, while the small white circle the third source. The latter is at a location consistent within the errors with that of 3FHL J1830.0-1518 (the radii indicate the 95\% confidence level error in the positions of the point sources).}
  \label{tsmap}
\end{figure}

We also made a TS map calculated after fitting a point source at the location of each pixel in the map. The LAT source associated to HESS J1825-137 was kept in the model used to obtain the map, which removes its emission. The results are shown in Fig. \ref{tsmap}. We found the positions of three point sources with the tool \emph{gtfindsrc}, a standard LAT tool that computes the TS over a grid of locations, and they are shown in the figure as well. The TS map shows significant gamma-ray emission in congruence with the residuals observed in Fig. \ref{fig1}.

The TS of the fit including the two sources with the highest significance (the green circles in Fig. \ref{tsmap}), with respect to the null hypothesis, is TS$_{\mbox{\tiny 2PS}}=80.4$. When adding a third source (the white circle in Fig. \ref{tsmap}), the result is TS$_{\mbox{\tiny 3PS}}=100.2$. The TS for the extended template and point sources cannot be compared using a likelihood-ratio test because the models are not nested. As an alternative, the Akaike information criterion can be estimated \citep[AIC,][]{1974ITAC...19..716A}. The AIC is defined as $2k -2\, \mbox{ln}L$, where $k$ is the number of parameters in the model. The best hypothesis is the one that minimizes the AIC. The three point-like sources have a total of 12 additional degrees of freedom while the semicircular template 6 (including its radius and orientation angle), with respect to the background model. The condition that the extended hypothesis is preferred with respect to the 3 point sources (AIC$_{\mbox{\tiny ext}} <\,$AIC$_{\mbox{\tiny 3PS}}$) translates to TS$_{\mbox{\tiny ext}}\,+ 12 >\,$TS$_{\mbox{\tiny 3PS}}$, which is the case. However, it remains possible that an extended source is actually the combination of point and extended sources that could be resolved in the future. It is interesting to compare the fitted spectra under the point source hypothesis. The most significant spots (with TS = 46 and 33) have spectral indices $1.43\pm0.23_{\mbox{\tiny stat}}$ and $1.78\pm0.25_{\mbox{\tiny stat}}$ and for the third location (with TS$\sim 20$), $1.7\pm0.4_{\mbox{\tiny stat}}$. Their best-fit coordinates are RA=$18^h30^m21.4^s$, Dec=$-16\degr00'40.3''$, RA=$18^h29^m36.0^s$, Dec=$-14\degr23'41.6''$, and RA=$18^h30^m10.6^s$, Dec=$-15\degr19'03.4''$ (J2000), respectively. The spectra and location of the third source found (the white circle in Fig. \ref{tsmap}) are consistent with those reported for the unidentified source 3FHL J1830.0-1518. Similarly hard spectra measured at spatially separated locations could indicate a common origin of the emission.

\subsection{Radial profile}
\begin{figure}
  \includegraphics[width=9cm,height=5.8cm]{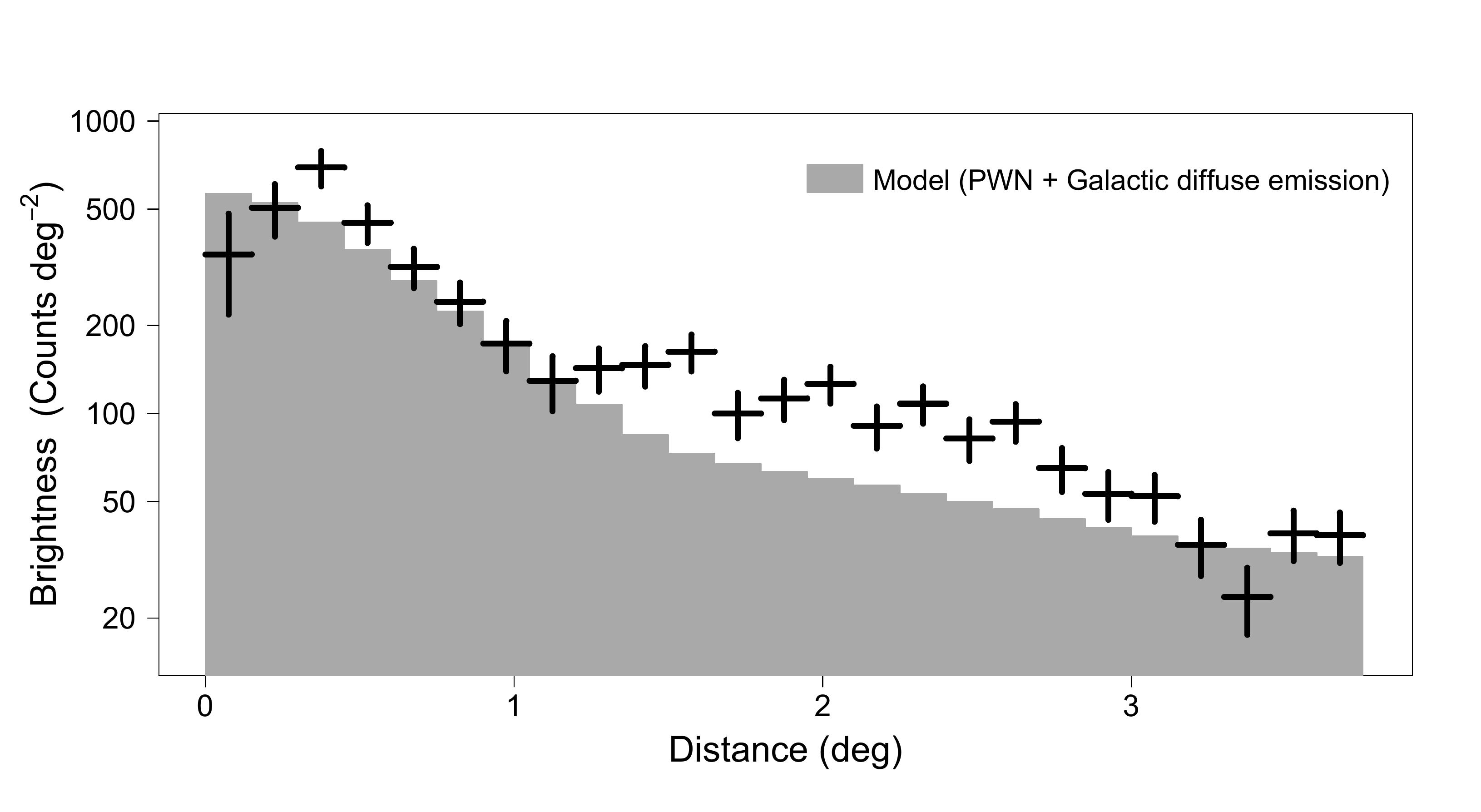}
  \caption{Radial profile of the LAT counts map and model obtained from the wedge shown in Fig. \ref{fig2} using front events and the P8R2\_CLEAN class above 10 GeV. The model components (gray bars) dominating the emission within the wedge were the 3FHL template for HESS J1825-137 and the diffuse Galactic emission. The source associated to LS 5039 was excluded and the origin of the profile is the center of the LAT template for HESS J1825-137. The number of counts is not corrected for exposure since this is relatively uniform across the region (the variations are $\leq 0.5$\%). No smoothing was applied. }
  \label{fig3}
\end{figure}

In order to further study the properties of the LAT emission we obtained a radial brightness profile from the data and best-fit model. The location and size of the wedge used to extract the profile are also shown in Fig. \ref{fig2}. We chose the origin to be the center of the LAT template for HESS J1825-137 in the 3FHL catalog. We excluded emission from the binary system LS 5039 with a $0.15\degr$-radius circular region around its reported location in the 3FHL catalog (the PSF 68\%-containment radius is $\leq 0.1 \degr$ above 10 GeV for these events). We determined the angular aperture of the wedge by first inspecting an azimuthal profile of the photons within 0.75$\degr$ and 3.0$\degr$ of the LAT centroid of the PWN using the same events. This allowed us to find the region where the emission deviates more significantly from the model. We found that a wedge with an angular aperture of $\sim 70\degr$ as shown in Fig. \ref{fig2} is appropriate to extract a radial profile, which is shown in Fig. \ref{fig3}. The profile of the data is in good agreement with the LAT model out to $\sim 1.2\degr$ from which point an excess is seen associated with the new region. We note that the model also underpredicts the emission around $0.4\degr$ from the center of the PWN, because the distribution of events deviates from a Gaussian morphology, with an enhancement of emission at this location.

\subsection{Spectral analysis}
We searched for spectral curvature in the emission by fitting different shapes using the template found in the previous steps. The resulting TS values for a log-parabola and a power-law with an exponential cutoff were 290.5 and 290.1, respectively. For one additional degree of freedom in the fit with respect to the simple power-law fit (with TS=281.5) these spectral shapes only produced a marginal improvement in the likelihood and we adopted the simple power-law ($\frac{dN}{dE} \propto E^{-\Gamma}$) as the best description of the spectrum.

\begin{figure}
  \includegraphics[width=9cm,height=5.8cm]{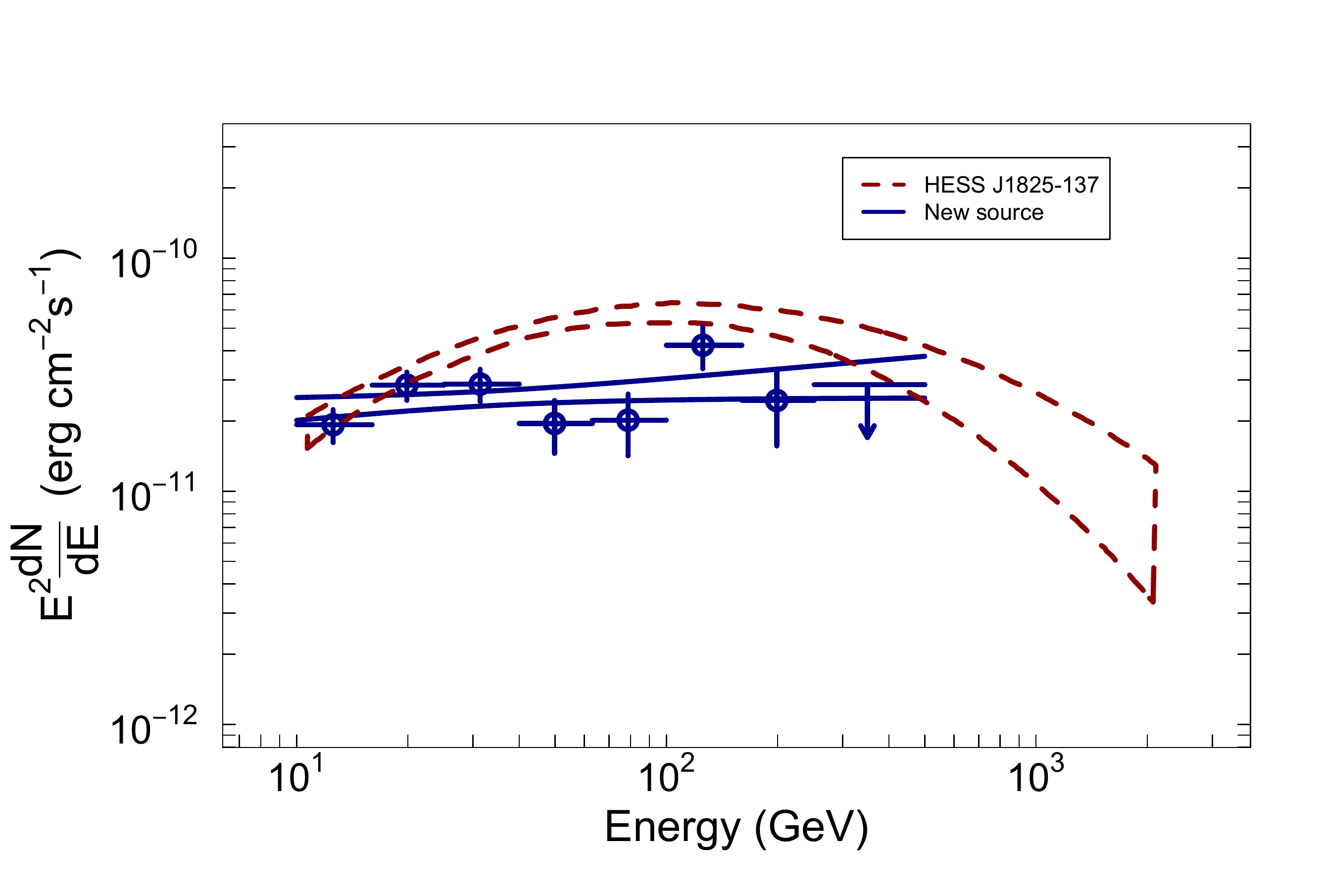}
  \caption{SED of the gamma-rays found to the south of HESS J1825-137. The blue lines represent the uncertainty band of the fit in the whole energy range and the points are calculated from the fluxes measured in each data bin using the best-fit template found. The red lines represent the uncertainty band of the published SED of HESS J1825-137 \citep[taken from][]{2017ApJ...843..139A} for comparison.}
  \label{sed}
\end{figure}

We also binned the LAT data in eight energy bins equally spaced logarithmically in the 10-500 GeV range to assess the significance of the emission in each bin. Gamma-rays from the region are seen in all bins except for the highest energy range, 250-500 GeV, where no significant emission is detected. In the 10-250 GeV range the resulting spectral index and integrated flux were $\Gamma = 1.92 \pm 0.07_{\mbox{\tiny stat}} \pm 0.05_{\mbox{\tiny sys}}$ and $(1.46 \pm 0.11_{\mbox{\tiny stat}} \pm 0.13_{\mbox{\tiny sys}})\times 10^{-9}$ photons cm$^{-2}$ s$^{-1}$. The spectral energy distribution (SED) obtained is shown in Fig. \ref{sed}, including both the fit model and the fluxes in each energy bin. For comparison, the SED of HESS J1825-137 above 10 GeV from the Fermi Galactic Extended Source Catalogue (FGES) is also shown \citep{2017ApJ...843..139A}.

In order to assess the influence of the emission from HESS J1825-137 that is not properly accounted for in the model (e.g., see Fig. \ref{fig3}) on our results, we repeated the search defining a new template which excluded the overlap region with the PWN. The highest TS value was obtained for the same template size and orientation as that described in section \ref{morphology}. The TS value was only slightly lower than before (278.2) and the resulting spectral parameters are fully consistent with the ones previously reported.

The estimated systematic errors combined two effects: the uncertainties in the effective area of the LAT and the uncertainty in the Galactic diffuse emission. The bracketing effective area method was applied as recommended by \cite{2012ApJS..203....4A} to propagate the instrument uncertainties onto the normalization and spectral index uncertainties. The errors due to the unperfect knowledge of the diffuse emission were estimated as in \cite{2016ApJS..224....8A}. The standard diffuse emission models used in most LAT analyses are optimized for studies of point sources and compact extended sources. The results obtained in this study were validated with a set of eight alternative models described by \cite{2012ApJ...750....3A} where the cosmic ray source distribution, the height of the cosmic ray halo and the HI spin temperature are varied. For all the alternative models the TS of the template representing the new emission region was greater than 218 and the estimated systematic errors were relatively small. Note that since the diffuse emission models available for LAT analyses are slightly dependent on the instrument response functions (IRFs) and our alternative models were developed for Pass 7 analysis the estimated errors are only approximate. Given the consistent results for the spectra found using these models and the standard model for Pass 8 analysis we believe that any systematics introduced by using different IRFs is not significant. This is particularly true in the $10-250$ GeV energy range where the effective area of the LAT is basically independent of energy. For example, the scaling factors applied to the standard Pass 7 emission model to obtain the standard model for Pass 8 analysis are in the range $0.98-1$ above 10 GeV\footnote{See \url{https://fermi.gsfc.nasa.gov/ssc/data/access/lat/Model\_details/Pass8\_rescaled\_model.html}}.

\section{CO data}\label{cloud}
In order to better understand the environment where the gamma-ray excess is located we studied observations of the molecular CO transition lines by \cite{2001ApJ...547..792D} which are used to indirectly detect molecular hydrogen. In all the available line velocities (measured with respect to the local standard of rest, $V_{lsr}$) in the data we saw that the region occupied by the gamma-rays is devoid of molecular clouds except for a small clump in the approximate range $V_{lsr} = 15-21$ km/s and another dimmer excess around 29 km/s. Adopting a standard galactic rotation model \citep{1993A&A...275...67B} these velocities correspond to near kinematic distances in the range of $1.6-2.2$ kpc and 2.8 kpc, respectively (the far distance solutions are in the range $13.4-14.6$ kpc). These results imply that the molecular material is not related to HESS J1825-137. Conversely, no CO emission was seen in the velocity range around the dispersion measure distance of the pulsar, $3.9\pm 0.4$ kpc.

\section{Discussion}
We found a considerably extended region of gamma-ray emission in the vicinity of the PWN HESS J1825-137 reaching $\sim 2.5\degr$ to the south of the position of the pulsar. In the 10-250 GeV range the spectrum of this emission is hard with a photon index $\sim 1.9$. At a distance of 3.9 kpc the corresponding luminosity is $1.5\times 10^{35}$ erg/s, or about 0.6 times the luminosity of HESS J1825-137 as determined from the same data set and the 3FHL template for this source. Several scenarios could explain the origin of these gamma-rays and we explore some of them in this section.

\subsection{High-energy particles from the PWN}
The emission from a PWN has a distinctive nonthermal signature produced by high-energy electrons and positrons (hereafter electrons) that is seen from radio to gamma-rays. The specific spectral and morphological features of the radiation depend on the environment and pulsar properties, such as its spin-down power \citep[see, e.g.,][]{2011MNRAS.410..381B}. The energy flux in the gamma-ray regime dominates at TeV energies but PWNe have been also detected by the \emph{Fermi} LAT at GeV energies \citep[e.g.,][]{2013ApJ...773...77A}. The emission is thought to be produced mainly by inverse Compton (IC) scattering of background low energy photons. The gamma-ray emission usually occupies a large region of space filled with particles accumulated over the age of the system. Electrons are believed to be transported by diffusion or convection to form the extended nebula. Observations and theoretical work show that likely a combination of both transport mechanisms is present \citep{2006ARA&A..44...17G,2012ApJ...752...83T,2016MNRAS.460.4135P}. The electrons responsible for the X-ray nebula have relatively low lifetimes due to cooling losses while TeV emitting electrons have lifetimes comparable to the characteristic age of the pulsar. The GeV and sub-TeV emission is mostly produced by relic uncooled electrons injected during earlier epochs of the nebula when the energy output from the pulsar was greater. It is then expected for the GeV gamma-rays to occupy a larger region of space or to be displaced from the TeV peak.

We consider only two energy loss mechanisms caused by the production of IC and synchrotron radiation \citep[although we neglect adiabatic losses, note that they might be important after a few tens of thousands of years for this system, as shown by][]{2011ApJ...742...62V}. Previous values found for the magnetic field in the PWN are in the range from $3$ to $10\,\mu$G \citep{2009PASJ...61S.189U,2011ApJ...742...62V,2011ApJ...738...42G}, and the field is expected to decrease farther from the pulsar \citep[e.g.,][]{2011ApJ...742...62V} down to typical ISM values of a few $\mu$G. We estimate that the radiative cooling time scale for electrons responsible for the GeV emission is larger than the age of the system. IC photons of energy $E_{\gamma}$ can be produced by electrons interacting with the cosmic microwave background (CMB) with typical energy $E_e \approx 18$ TeV $(E_{\gamma}/1\,\mbox{TeV})^{1/2}$ in the Thomson regime \citep[e.g.,][]{2011ApJ...742...62V}. Thus, photons in the range $10-500$ GeV are mostly produced by electrons of energies in the range $2-12$ TeV. For these electrons the IC cooling time scale in the CMB is $\tau_{IC} \approx 140\, (E_e/10 \,\mbox{TeV})^{-1}$ kyr, which is quite large. On the other hand, the lifetime of electrons due to synchrotron cooling in a magnetic field $B$ is given by $\tau_{sync} \approx 8200\, (E_e/10\,\mbox{TeV})^{-1}(B/10\, \mu\mbox{G})^{-2}$ yr. For a mean field of $3\, \mu$G the cooling time scale of 10 TeV electrons is 90 kyr which is around twice the value given by some estimates of the actual age of the system \citep[e.g.,][]{2011ApJ...742...62V}.

If, as we have shown, it is possible for the electrons in the region to survive for long enough, we can estimate the required advection speeds and diffusion parameters to account for the observed extension of the GeV emission, leaving a more detailed modeling for the future. Fig. \ref{fig1} shows emission extending $\sim 2.5\degr$ from the pulsar and this emission seems to occupy a volume that is limited by the SNR shock front. At a distance of 3.9 kpc this corresponds to 170 pc. The most extreme speed would be obtained for a propagation time equal to the characteristic age of the pulsar, or 7900 km/s, and 4000 km/s for an age of 40 kyr for the system. These speeds are factors of 1.9 to 0.95 the speed resulting from an extrapolation of the H.E.S.S. velocity profile to 170 pc \citep{2011ApJ...742...62V}.

On the other hand, in a scenario where diffusion accounts for the propagation of leptons to large distances away from HESS J1825-137, considering the simple case of energy-independent diffusion in a scale of $R=170$ pc and two time periods of $t=21$ kyr and 40 kyr, the required diffusion coefficients are $D=R^2/2t = 2\times 10^{29}$ cm$^2$ s$^{-1}$ and $1.1\times 10^{29}$ cm$^2$ s$^{-1}$. These are factors of 4 and 2, respectively, the values found by \cite{2009PASJ...61S.189U} in the inner nebula, and similar to those predicted for the ISM in a standard model \citep[e.g.,][]{2007ARNPS..57..285S}.

\begin{figure}
  \includegraphics[width=9.7cm,height=8cm]{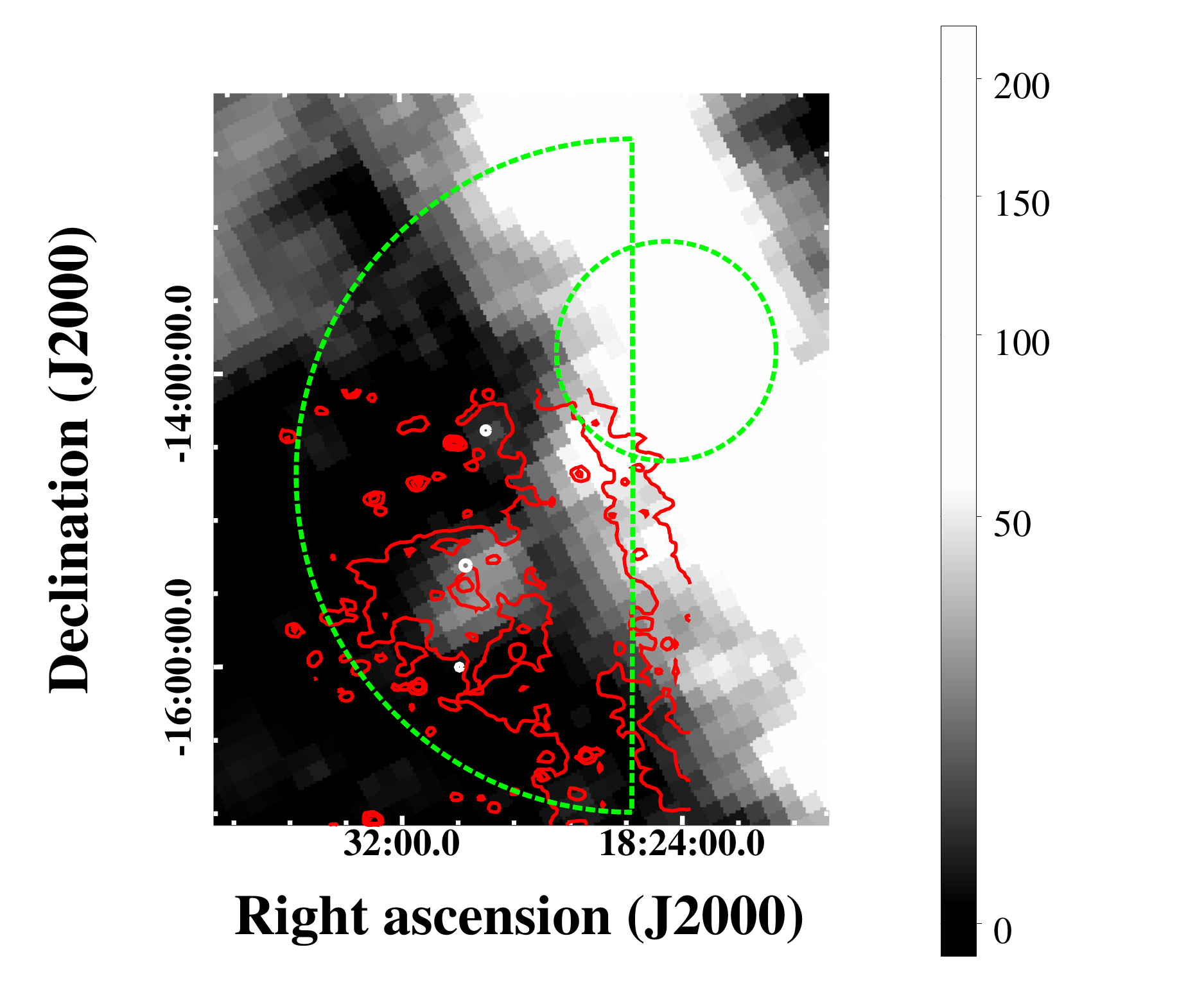}
  \caption{Integrated J=1 to 0 CO emission map \citep[in K Km/s,][]{2001ApJ...547..792D} showing the same regions from Fig. \ref{tsmap} and the infrared emission contours from a 12 $\mu$m observation \citep[at the levels of 6, 12 and 18 MJy/sr, from][]{2005ApJS..157..302M}. The cloud discussed in the text is seen at the center of the image.}
  \label{fig4}
\end{figure}

An interesting feature in the LAT data is the radial profile shown in Fig. \ref{fig3}. The shape of this profile indicates that there is either and enhancement in the electron density in this region or some source of low-energy photons that are up-scattered by the electrons that were injected by the HESS J1825-137 system. Observations from the Infrared Astronomical Satellite \cite[IRAS,][]{2005ApJS..157..302M} in the wavelength range 12-100 $\mu$m show the presence of emission that we believe based on its spatial correspondence is associated to the CO cloud discussed above. However, we determined from our kinematic distance estimates that this cloud is not related to HESS J1825-137. Therefore the infrared emission is likely not responsible for the GeV excess. Fig. \ref{fig4} shows the locations of the CO and infrared emission.

It might be possible to achieve such a radial profile from the accumulation of old electrons through the history of the system. Interestingly, the model by \cite{2011ApJ...742...62V} predicts a flat brightness profile in the LAT energy regime within $1\degr$ from the pulsar.

From the point of view of the energetics the pulsar origin of the GeV gamma-rays in the southern region is not problematic. We saw that the $10-250$ GeV luminosity is about half of the PWN luminosity measured by the LAT in the same energy range.

Finally, we might speculate that the gamma-rays are produced by electrons that escaped from the PWN, were re-accelerated through interactions with the reverse shock and diffused to their current location, or reached the observed region through a combination of the above-mentioned transport mechanisms.

\subsection{Gamma-rays from particles accelerated in the host SNR}
H$\alpha$ emission with similar properties to those of SNR shock fronts is seen $\sim 1.8\degr$ to the east of PSR J1826-1334. As shown by \cite{2018ApJ...860...59K} there is a reasonable scenario capable of explaining the growth of the SNR to the corresponding size at the distance to the pulsar. The location of this H$\alpha$ emission and other rim-like emission to the south with respect to the gamma-ray region found here is seen in Fig. \ref{fig1}. The gamma-ray morphology of middle-aged SNRs expanding in a uniform medium is expected to be shell-like \citep[e.g.,][]{2016ApJS..227...28T}. This is the case of the Cygnus Loop SNR, for example, whose GeV emission shows a correlation with the H$\alpha$ emission and the X-ray rims \citep{2011ApJ...741...44K}. Similarly to features seen in the Cygnus Loop and other SNRs there is an overlap of the H$\alpha$ emission to the south of HESS J1825-137 and the gamma-rays, but this is not the case with the H$\alpha$ found towards the east of HESS J1825-137. Furthermore, the origin of the H$\alpha$ emission in the south is not known and it could be the case that it is not produced by a shock front. A shell-like morphology of the GeV emission is not evident from Fig. \ref{fig1} and more detailed studies will be required to better characterize the emission. However, from the point of view of the energetics, an association of the GeV emission with particles accelerated in an SNR is not problematic. The gamma-ray emission has a luminosity that is typical of SNRs (if located at the distance to HESS J1825-137). The spectrum of the emission is hard and we could speculate that the emission comes from TeV particles once accelerated at the front or reverse shock of the host SNR and remained confined within the shell. The GeV spectrum is similar to that of other extended shell or shell candidates such as G150.3+4.5 and G350.6-4.7 \citep{2017ApJS..232...18A,2018MNRAS.474..102A}.

\subsection{A source (or sources) unrelated to HESS J1825-137}
Other objects could be responsible for the GeV emission seen to the south and east of HESS J1825-137. For example, it is believed that star-forming regions (SFRs) can accelerate cosmic rays through the combined winds of massive stars. Several of these regions have been detected by the LAT as extended sources with hard spectra \citep[with photon spectral indices $2.1-2.2$, e.g.,][]{2011Sci...334.1103A,2017ApJ...839..129K}. A catalog of SFRs by \cite{2002ARep...46..193A} shows several objects seen in the direction of the cloud discussed in Section \ref{cloud}, which is unrelated to HESS J1825-137. If located at a distance of $\sim 1.9$ kpc the resulting $0.2-500$ GeV luminosity (extrapolating the observed spectrum) is $8\times 10^{34}$ erg/s, which is similar to the luminosity of other SFRs such as the Cygnus Cocoon \citep{2017ApJ...839..129K}. From the point of view of the energy budget this scenario is plausible. A confirmation of the existence of OB associations and the corresponding bubbles in the region is necessary to back this possibility. No such structure is evident from infrared observations \citep[e.g.,][]{2005ApJS..157..302M}.

Several compact shell SNRs (with radii in the range $0.14-0.28 \degr$) are known in the region of the gamma-ray emission: G16.2-2.7, G17.4-2.3, G17.8-2.6, G18.9-1.1, \citep{2009ApJ...694L..16H,1997AJ....114.2058G}. Even though a single SNR would be capable of accounting for the observed gamma-ray luminosity, if these shells were accelerating particles the LAT images would show compact and distinct sources given the resolution of the LAT above 10 GeV. Indeed the fitted spectra of point sources at the locations with greater excesses are similar to each other with hard photon indices, which is evidence for a common origin of the emission.

\section{Summary}
We have found a significantly extended region of gamma-ray emission in the vicinity of the PWN HESS J1825-137 with a hard spectrum above 10 GeV. The emission seems to be bounded by the candidate SNR shock front seen in H$\alpha$ images. If located at the same distance to HESS J1825-137 the luminosity of the gamma-rays is about half of that of the PWN itself in the same energy range. The parameters found from simple estimates of particle diffusion or convection from the PWN to explain the extension of the source are reasonable, although a detailed model has to be applied in the future to gain more insight on the physical processes playing a role in the growth of the candidate electron nebula. Other previously unknown sources such as a SNR or a star-forming region could explain the spectral and morphological properties of the radiation, and multiwavelength observations should be carried out in the future to find the counterpart of the GeV gamma-rays.

\section*{Acknowledgements}

We thank J.A. Hinton for useful discussions and the anonymous referee for the valuable comments made to improve the clarity of the work. This project has received funding from the European Union's Horizon 2020 research and innovation programme under the Marie Sk\l{}odowska-Curie grant agreement No 690575. Financial support from Universidad de Costa Rica is also acknowledged.

\bibliographystyle{mnras}
\bibliography{references}

\bsp	
\label{lastpage}

\end{document}